\documentclass[12pt]{iopart}

\usepackage{amsmath,amssymb}
\usepackage{amsfonts} 
\usepackage{iopams}
\usepackage[dvips]{graphicx}
\usepackage[spanish, english]{babel}
\DeclareFixedFont{\zcal}{T1}{pzc}{m}{it}{14pt}

\newcommand{\IntSum}{\ensuremath{\sum  \!\!\!\!\!\!\!\! \int }} 

\newcounter{listado}

\begin{document}
\bibliographystyle{unsrt}

\title[JPB]{Time-resolved resonant photoionization of He using a time-dependent Feshbach method
with ultrashort laser pulses.}

\author{C. M. Granados-Castro and J. L. Sanz-Vicario}

\address{Grupo de F\'{\i}sica At\'omica y Molecular. Instituto de F\'{\i}sica \\
 Universidad de Antioquia, Medell\'{\i}n, Colombia.}
\ead{sanjose@fisica.udea.edu.co}

\begin{abstract}
We study the photoionization and autoionization of Helium atom subject to ultrashort laser pulses by using a
Feshbach formalism in the time domain. We solve the time-dependent Schr\"odinger equation in terms of a 
configuration interaction (CI) spectral method, in which the total wavefunction  is expanded with configurations
defined within bound-like ($\mathcal{Q}$) and scattering-like ($\mathcal{P}$) halfspaces. The method allows
one to provide accurate descriptions of both the atomic structure (energy positions and widths) and 
the photodynamics. We illustrate our approach by i) calculating the time-resolved one-photon ionization
below the He$^+$ ($n$=2) ionization threshold, from $1^1S^e$ and $2 ^1P^o$ initial states, then reaching
the lowest autoionizing states of $^1S^e$, $^1P^o$ and $^1D^e$ final symmetries ii) studing the temporal
formation of the Fano profile of $^1P^o$ resonances and iii) showing its performance in obtaining the
perturbative long-time limit of one- and two-photon ionization cross sections using ultrashort laser pulses 
following a recently developed procedure in Phys. Rev. A, {\bf 77}, 032716 (2008).
\end{abstract}

\submitto{\JPB}

\section{Introduction}

In the last decade the experimental development of high intensity, high frequency ultrashort laser pulses down to the
attosecond time scale has led to the investigation of electron photodynamics in its natural time scale \cite{Drescher:2002,Krausz:2009}. In general,
the temporal behaviour of many-electron atoms subject to a strong laser field still remains a challenging theoretical problem when
dealing with multiple excitation and ionization, including Auger phenomena, which requires to account for the appropriate representation
of the electron correlation (responsible for the autoionization) along with the solution of the time-dependent Schr\"odinger equation.
For instance, it is interesting to study the coherent excitation and decay of metastable superexcited states using pump-probe laser pulses 
with time delays comparable in magnitude to autoionization lifetimes in order to trace the fast dynamics involved. The decay dynamics of doubly excited states in the Helium atom
has already been addressed using time-dependent close-coupling methods implemented in a numerical lattice \cite{Hu:2005}. Also, 
attosecond pump probe laser schemes have been proposed to probing ultrafast electron motion in singly excited \cite{Hu:2006}
and doubly excited states in Helium \cite{Argenti:2010}. In particular, the latter work is based on a sophisticated multichannel close-coupling method 
using B-splines with very large radial boxes, large $L$-values in the angular momentum expansion, introduction of absorbing potentials, etc., 
ingredients that require in general a high computational effort. 

Considering fast evolving phenomena, it has been of recent interest the time-resolved formation of Fano profiles in the atomic photoionization spectra, using both simplified models and
{\em ab initio} methods \cite{Wickenhauser:2005,Mercouris:2007,Chu:2010}. In this respect, experimental attosecond resolution of the He autoionization process has been recently reported  \cite{Gilbertson:2010}, demonstrating control over the two interfering paths, direct ionization and autoionization, that shape the profile of the Fano resonance \cite{Fano:1961} (for a recent study on the modification of the asymmetry in
the Fano profiles with ultrashort pulses see also \cite{Bengtsson:2012}). New emergent phenomena arising in the autoionization dynamics of laser-dressed atoms (pump probe schemes) where excited resonances are coupled by an IR field, have been mostly studied 
using simplified models \cite{Chu:2011,Chu:2012}, that although reproduce the relevant features of experiments and support the physical explanation, they may prove insufficient in other scenarios that require fully {\em ab initio} methods \cite{arxiv:2012}.

In this work we describe a time dependent Feshbach method applied to atomic photoionization of He subject to ultrashort laser pulses.
The time independent Feshbach formalism has been widely applied in atomic physics in the last decades but the time dependent formalism 
in the present form has been used only in the molecular context recently \cite{Sanz-Vicario:2006}. A similar time-dependent approach (but not identical in the form of the wave packet expansion and the required final projections) was proposed to study the resonant and non resonant ionization of He by XUV intense ultrashort laser pulses \cite{Hasbani:2000}.  In particular, we describe here some practical details to 
generate discretized non resonant continuum states for a given selected energy from the solution of the eigenvalue problem. 
To gauge the performance of the method, we provide simple illustrations like the computation of one- and two-photon ionization cross section 
in He.  For weak fields, the dependence of the one-photon ionization probability with the pulse duration obeys only to spectral effects. In fact,
the temporal dependence in the transition amplitudes calculated in time dependent first order perturbation theory  is exactly factored out in 
one-photon absorption processes and approximately in the multiphoton case. This property allows to obtain cw
perturbative photoionization cross sections from amplitudes extracted over the range of continuum energies within the spectral bandwidth
of short laser pulses.  We also study the formation of the profile of the Fano resonant peak in the time domain using ultrashort fs pulses, showing
transient oscillations due to the two-path interference, which eventually vanish to yield the asymptotic stationary result. 

Atomic units are used unless otherwise stated.

\section{Theory}

The Feshbach projection method \cite{Feshbach:1958,Feshbach:1962} has shown to be a powerful method
to describe resonance phenomena in scattering processes. Its application to the atomic electronic structure
can be found elsewhere (\cite{Temkin:1985a} and references therein) although its practical implementation
is mostly reduced to atomic systems with two and three electrons. A detailed study of the application of the 
stationary Feshbach method in He has been performed by S\'anchez {\em et al} \cite{Sanchez:1995a}. Also, 
after the pioneering work of Temkin and Bathia on three-electron systems \cite{Temkin:1985b}, the Feshbach formalism has been 
recently revisited and applied to the Li atom in our laboratory \cite{Cardona:2010},
including and assessing all the required ingredients of the rigorous formalism. 
The Feshbach projection operator formalism is based
on the introduction of projection operators $\mathcal{P}$ and $\mathcal{Q}$, satisfying completeness
($\mathcal{P}$+$\mathcal{Q}$=1), idempotency ($\mathcal{P}^2$=$\mathcal{P}$, $\mathcal{Q}^2$=$\mathcal{Q}$)
and orthogonality ($\mathcal{P}\mathcal{Q}$=$\mathcal{Q}\mathcal{P}$=0), which project the total wavefunction 
onto nonresonant scattering-like and bound-like halfspaces, respectively.  These projected wave functions must
also satisfy the asymptotic boundary conditions $lim_{r_i \to \infty} \mathcal{P} \Psi$=$\Psi$ and $\lim_{r_i \to \infty}
\mathcal{Q} \Psi$=0, the latter indicating the confined nature of the localized part of the resonance.

By introducing the splitting of the total wave function $\Psi$=$\mathcal{Q}\Psi+\mathcal{P}\Psi$ into the time
independent Schr\"odinger equation $H \Psi$=$E\Psi$, working equations for the bound-like and the {\em non-resonant} scattering-like 
parts arise as follows (see, for instance,  \cite{Cardona:2010}):
\begin{subequations}\label{eq:QPspace}
\begin{eqnarray}
(\mathcal{Q}H\mathcal{Q}-\mathcal{E}_n ) \mathcal{Q}\Phi_n=0 \label{eq:Qspace}\\ 
(  \mathcal{P} H'  \mathcal{P}-E)  \mathcal{P} \Psi^{0} = 0 \label{eq:Pspace},
\end{eqnarray}
\end{subequations}
where $H'$ is the operator containing the atomic Hamiltonian plus an optical potential devoid of any resonant contribution 
from the state $\mathcal{Q}\Phi_s$ with energy $\mathcal{E}_s$, i.e., $H'$=$H+ V^{n\ne s}_{opt}$ where
\begin{equation}
V^{n\ne s}_{opt}=\IntSum_{n \ne s} \mathcal{P}H\mathcal{Q} \frac{|\Phi_n\rangle \langle \Phi_n|}{E-\mathcal{E}_n} \mathcal{Q}H\mathcal{P}.
\end{equation}

It is worth noting that the Hamiltonian splits into 
$H$=$\mathcal{Q}H\mathcal{Q}$+$\mathcal{P}H\mathcal{P}$+$\mathcal{Q}H\mathcal{P}$+$\mathcal{P}H\mathcal{Q}$, where the last two terms are responsible for the coupling between both halfspaces which ultimately causes the resonant decay into the continuum. In practice one starts by solving Eq. \eqref{eq:Qspace} for the $\mathcal{Q}$ space
with a configuration interaction method to obtain a first approximation to the location of resonant states and it implies to use $\mathcal{Q}$=1-$
\mathcal{P}$, where $\mathcal{P}$=
$\mathcal{P}_1$+$\mathcal{P}_2$-$\mathcal{P}_1\mathcal{P}_2$ with $\mathcal{P}_i$ being a one-particle projection operator. In this
work we are restricted to doubly excited states lying below the second ionization threshold of the He atom, so that $\mathcal{P}_i =| \phi_{1s} 
({\bf r}_i) \rangle \langle \phi_{1s} ({\bf r}_i)|$. Therefore, the $\mathcal{Q}$ operator removes all those configurations containing the $1s$ orbital,
then avoiding the variational collapse to the ground state ($1s^2$), to singly excited states ($1s n\ell$) and removing also
the single ionization continuum ($1s\epsilon \ell$). Instead of diagonalizing the $\mathcal{Q}H\mathcal{Q}$ problem, one may solve the
equivalent but simpler eigenvalue problem involving an effective Hamiltonian that consist of the full Hamiltonian and a Phillips-Kleinman 
pseudopotential \cite{Martin:1987}, $H_{eff}$=$H$+$M\mathcal{P}$,
where $M$ is a large real number. The effect of $M\mathcal{P}$ is to project upward in energy all eigenstates associated with
the $\mathcal{P}$ halfspace, so that the lowest variational eigenstates correspond to the Rydberg series of resonant discrete states below the
corresponding ionization threshold.

To build the nonresonant electronic continuum we solve Eq. (\ref{eq:Pspace}) for the $\mathcal{P}$ subspace using a basis of
two-electron configurations $\{\varpi_k\}_{k=1}^N$, with $\varpi_k( {\bf x}_1,{\bf x}_2) = \mathcal{A} \left( \phi_{1s}({\bf x}_{1}) \phi_{k \ell}({\bf x}_2) 
  \right )$, where we use the notation ${\bf x}_k $=$(r_k,\Omega_k,s_k)$ (radial $r_k$, angular $\Omega_k$, and spin $s_k$ coordinates).
This conforms a {\em static-exchange approximation} for the {\em nonresonant} continuum wave function $\mathcal{P}\Psi^{0}$, which is orthogonal 
to $\mathcal{Q}$, as proved by construction, since $\phi_{1s}$ is used to build the $\mathcal{P}$ projection operator. As a general rule, the ground 
state, the singly  and doubly excited states are obtained with a configuration interaction method, using a truncated set of antisymmetrized two-electron configurations 
$\{ \varpi_n^{L,M_L,S,M_s} \}_{n=1}^{N}$ adapted to the $L,S$ total angular momenta and  generated with products of atomic orbitals $\phi_{n\ell}$, as  follows
\begin{equation}
\varpi_n^{L,M_L,S,M_S} ({\bf x}_1,{\bf x}_2)  = \mathcal{A} \left( \phi_{n_1 \ell_1} (r_1) \phi_{n_2 \ell_2} (r_2)  \mathcal{Y}_{\ell_1,\ell_2}^{L,M_L}
(\Omega_1,\Omega_2) \chi^{S,M_s} (s_1,s_2) \right),
\end{equation}
where $ \mathcal{Y}_{\ell_1,\ell_2}^{L,M_L}$ is the bipolar spherical harmonics, $\chi$ the spin wave function, and $\mathcal{A}$ corresponds to 
the antisymmetrizing operator. In the stationary implementation of the Feshbach formalism, one may readily compute the energy shift correction
$\Delta_s$ for the resonant state $\mathcal{Q}\Phi_s$ due to the perturbation of the surrounding continuum with
\begin{equation} \label{eq:shift}
\Delta_s = \IntSum_{E' \ne E_s} dE' \frac{|\langle \Phi_s | \mathcal{Q}H\mathcal{P}| \mathcal{P}  \Psi^{0} (E') \rangle|^2}{E_s-E'}, 
\end{equation}
so that its actual energy is $E_s=\mathcal{E}_s + \Delta_s$. $\mathcal{Q}H\mathcal{P}$ couplings also allow to compute the resonant widths
using a Fermi's golden rule,
\begin{equation} \label{eq:width}
\Gamma_s= 2\pi | \langle \Phi_s |\mathcal{Q}H\mathcal{P} | \mathcal{P}\Psi^{0} (E=E_s)|^2 
\end{equation}
as well as the $q$-Fano profile parameter \cite{Sanchez:1995a}. We have described above the basic ingredients to perform stationary calculations within the Feshbach formalism. We now proceed to describe our implementation of the time-dependent Feshbach method to be applied to atomic photoionization phenomena with ultrashort laser pulses.

\subsection{Time dependent Feshbach method.}

The time dependent Schr\"odinger equation (TDSE) for a two-electron atomic system exposed to a laser field in the dipolar approximation
reads
\begin{equation} \label{eq:tdse}
\left(  H + V(t) -i \frac{\partial }{\partial t }  \right) \Psi ( {\bf x}_1, {\bf x}_2,t ) = 0
\end{equation}
where $V(t)$=$( {\bf p}_1$+${\bf p}_2)$$\cdot$${\bf A}(t)$ ($V(t)$=$({\bf r}_1$+${\bf r}_2)$$\cdot$${\bf E}(t)$) corresponds to the laser-atom interaction in the velocity
(length) gauge. The vector potential for a $z$-linearly polarized laser pulse is taken as ${\bf A}(t)$=${\bf \hat{e}_z} f(t) cos(\omega (t-T/2))$, where
$f(t)$ is a shape function for the pulse envelope taken here as $f(t)$=$\sin^2 (\pi t/T)$ and it is defined in the time interval $t\in[0,T]$
and zero elsewhere. The field amplitude is related to the laser intensity through $I$[W/cm$^2$] = 3.5095$\times$10$^{16} E^2_0$ [a.u.] and to
the vector potential amplitude with $A_0=E_0/\omega$.

We make use of a spectral method by expanding the time-dependent wavefunction  with the stationary 
Feshbach eigenstates of the $\mathcal{Q}H\mathcal{Q}$ and $\mathcal{P}H\mathcal{P}$ projected Hamiltonians. This means that our expansion is 
based on the asymptotic Hamiltonian, since $\mathcal{Q}H\mathcal{P}$ couplings vanish not only for $r_i \to \infty$, but they are also not effective 
for $t \to \infty$ since eventually all the $\mathcal{Q}$ resonant population leaks to the continuum $\mathcal{P}$ halfspace. To work with an asymptotic basis of eigenstates has the advantage of avoiding projections of a propagated full wavepacket (without Feshbach partitioning) and therefore the calculated expansion coefficients $C_i(t)$ for $t \gg \{T,1/\Gamma_s\}$ correspond to physical amplitudes
by themselves.  Then our expansion reads   
\begin{eqnarray} \label{eq:ansatz}
\Psi( {\bf x}_1,{\bf x_2} ) & =  \sum_{b} C_{b} (t) \Xi_{b} ({\bf x}_1, {\bf x}_2) e^{-i E_{b} t} \nonumber \\
                                          & +   \sum_{r}  C_{r} (t) \mathcal{Q}\Phi_{r} ({\bf x}_1, {\bf x}_2) e^{-i E_{r} t} \nonumber \\ 
                                          & +   \int dE  C_{E} (t) \mathcal{P}\Psi^{0}_{E} ({\bf x}_1, {\bf x}_2) e^{-i E  t} 
\end{eqnarray}
where $\{ \Xi_b, E_b \}_{b=1}^{N_b}$ correspond to the set of bound states (ground plus singly excited states), $\{ \mathcal{Q}\Phi_r, E_r \}_{r=1}^
{N_r}$ to the
resonant doubly excited states and $\{\mathcal{P}\Psi^0_E, E\}_E$ refer to the nonresonant continuum states. Although rigorously the
bound states pertain to the $\mathcal{P}$ space, in our implementation they are calculated separately as eigenstates of the full Hamiltonian
$H$. Indeed, the CI basis of the static exchange approximation used to expand the $\mathcal{P}$ halfspace provides rather poor results 
for the bound states. Using $\Xi_b$ states introduces non-orthogonalities in the basis set of configurations among states with 
the same {$L,S,\pi$} symmetries. For one-photon absorption ($L\to L\pm1,S,\pi \to \pi'$), the lack of orthogonality is irrelevant, but
for multiphoton processes it must be checked out carefully, introducing the corresponding overlaps in the couplings described below.
Now introducing the ansatz of Eq. (\ref{eq:ansatz}) in the TDSE (\ref{eq:tdse}) one arrives to a set of  coupled differential equations in the 
interaction picture, which may be written in packed block form, with $n$=$\{b,r,E\}$, as follows 
\begin{equation} \label{eq:coupled}
i \begin{pmatrix} 
\dot{ \mathbf{C}_b}  \\ \dot{\mathbf{C}}_r \\ \dot{\mathbf{C}}_E 
\end{pmatrix}=
e^{i E_n t}
\begin{pmatrix}
0                & V(t)_{b,r} & V(t)_{b,E} \\
V(t)_{r,b}  &   0             &  \mathcal{Q}H\mathcal{P}_{r,E} \\
V(t)_{E',g} &  \mathcal{P}H\mathcal{Q}_{E',r} & V(t)_{E',E}
\end{pmatrix}
e^{-iE_m t} 
\begin{pmatrix}
\dot{ \mathbf{C}}_b \\ \dot{\mathbf{C}}_r \\ \dot{\mathbf{C}}_E
\end{pmatrix},
\end{equation}
where $V(t)_{n,m}$ are coupling matrix elements corresponding to the laser interaction and $\mathcal{Q}H\mathcal{P}$ are 
electrostatic couplings since neither $\mathcal{Q}\Phi$ nor $\mathcal{P}\Psi^0$ are eigenstates of the field free Hamiltonian.
The latter couplings do not depend on time and therefore, once the laser pulse ends, they are still active to {\em empty} the resonant 
population until its full depletion. The set of coupled equations is solved subject to an initial condition $C_n$($t$=0)=$\delta_{n 0}$. 
In conclusion, the time-dependent Feshbach method relies on the preliminar calculation of the stationary eigenstates 
($\mathcal{Q}H\mathcal{Q}$ and $\mathcal{P}H\mathcal{P}$) and couplings $\mathcal{Q}H\mathcal{P}$ of the projected Hamiltonian, along with the dipolar couplings among them.

In other works (see, for instance, \cite{Huens:1993,Scrinzi:1998,Grosges:1999,Lagmago:2001}) the total wavefunction is expanded in terms
of the eigenfunctions of the complex-rotated (non-hermitian) unperturbed Hamiltonian $H (r e^{i\theta})$. Due to the presence
of a negative imaginary part in the eigenenergies of both the resonant and the continuum states, the time propagation allows for
the decay of the metastable state as well as to avoid non-physical reflections of the wave packet at the boundary box. Also, in the 
complex scaling (CS) method the complex eigenstate corresponding to the resonant pole includes both the bound-like and the scattering-like 
part of  the resonant state and Fano profiles in photoionization spectra may be obtained straightforwardly with a reduced number of 
final complex eigenstates since each single continuum pole represents indeed a bunch of unrotated continuum states, i.e., takes into account the 
presence of neighbouring continuum levels. The description
in terms of the Feshbach formalism requires instead a higher density of continuum states. The Feshbach approach treats separately
the bound-like and the scattering-like parts of resonances, allowing for the population transfer between both halfspaces through the 
specific inclusion of the electrostatic couplings $\mathcal{Q}H\mathcal{P}$. Concerning the
laser-atom interaction, the dipolar operator in the CS method involves the simultaneous coupling of both the bound and the scattering part 
of the resonance with other states, let them be bound, continuum or another resonant state. At variance, in the Feshbach method the dipolar 
couplings are performed separately between wave functions located in $\mathcal{P}$ and/or $\mathcal{Q}$ halfspaces.

Finally, multiphoton ionization cross sections can be obtained in the time domain using weak laser pulses with finite (but long enough) time 
duration (to be compared with time-independent perturbative results in the energy domain) using the expression (see, for instance, 
\cite{Sanz-Vicario:2006})

\begin{equation} \label{eq:crosssection}
\sigma [cm^{2N} s^{N-1} ] = \left(    \frac{\omega [Joules]}{I [W/cm^2]} \right)^N  \frac{P_{TDSE}}{C(N) [s]}
\end{equation}
where $\omega$ is the central frequency of the absorbed laser pulse, $I$ the laser intensity and  $C(N)$=$\int_{0}^T dt f(t)^{2N}$
 takes into account the effective pulse duration using its envelope $f(t)$ and it consists of a numerical factor times the pulse duration,
 i.e., $C(1)$=3/8$T$, $C(2)$=35/128$T$, and so on.  The total ionization probability is directly obtained from the coefficients $C_E$
 in the numerical solution of the coupled equations (\ref{eq:coupled}) with $P_{TDSE}$=$\sum_{E_i} |C_{E_i} (t>T)|^2$, i.e., just 
 a summation over all discretized continuum states (a simple quadrature of the integral) \cite{Reading:1979}.

\section{Computational details}

\subsection{Bound states.}
Our CI method is based on the expansion in terms of antisymmetrized products of atomic orbitals, the latter expanded in B-splines
polynomials enclosed within a finite box of length $L$. B-splines have been widely used in the last years and for a fuller description
the reader is referred to \cite{Bachau:2001}. A very precise ground state energy for He atom can be obtained with B-splines basis using
an exponential knot sequence and 40 B-splines with order $k$=7, generating one-electron orbitals with $\ell \le 3$, with a full 
CI wave function of 3280 configurations, which yields the energy -2.903321 a.u. to be compared with that of Pekeris \cite{Pekeris:1958},
-2.903742 a.u.  Nevertheless, since one is mainly interested in processes where the electronic continuum is mostly involved, the exponential
knot sequence is not well adapted to the description of continuum wave functions, due to the poor electronic density in the energy region
of interest. Instead, we have used in this work an exponential-linear sequence of knot-points within a box of length $L$=150 a.u., with 200 B-splines, 
in such a way that the sequence is exponential below 15 a.u. and linear in the rest of the box. This allows for a good description of the oscillations of 
the continuum wave functions in the full range of the box. Therefore, 27 bound states of symmetry $^1S^e$ (10 states), $^1P^o$ (9) and $^1D^e$ 
(8) are obtained with 8230, 9248, and 10986 configurations, respectively, using two-electron angular configurations 
built with orbitals with $\ell \le 3$. Eventually, for the $^1S^e$ ground state, the basis of orbitals may be supplemented with Slater-type orbitals
expanded in terms of B-splines to improve the convergence of the ground state energy \cite{Sanchez:1995a,Sanchez:1995b}.

\subsection{$\mathcal{Q}$-subspace.}
For the $\mathcal{Q}H\mathcal{Q}$ resonant space, we perform CI calculations with the same configurational basis set but removing the
$1s$ orbital. Then we are able to obtain 19 doubly excited states of symmetry $^1S^e$,  26 states for $^1P^o$, and 25 states for $^1D^e$, using
8456, 9135 and 10861 configurations, respectively. We do not intend here to produce benchmark results, since we have not followed 
a systematic optimization procedure, but to produce fairly good results to show the performance of the time-dependent Feshbach method.
To illustrate the accuracy of our computation, we report  in table \ref{TableI} the energies (including the energy shift correction of Eq. (\ref{eq:shift})) 
and widths (or lifetimes) for the $^1S^e$, $^1P^o$ and $^1D^e$
lowest doubly excited states below the $N$=2 threshold, and compared to accurate results obtained by Chen \cite{Chen:1997} using the
saddle-point complex rotation method. For a more comprehensive comparison, the reader is referred to \cite{Chen:1997}.

\begin{table}[htb]
\caption{\label{TableI} Energy positions (in a.u.), witdhs (in a.u.) and lifetimes (in fs unless otherwise indicated) for the lowest four $^1S^e$, $^1P^o$ 
and $^1D^e$ 
doubly excited states in He $^1S^e$ below the $N$=2 threshold. The values of the present work are compared with the saddle-point complex-
rotation results of 
Ref. \cite{Chen:1997}. The notation a[b] indicates a$\times$10$^b$.} 
\begin{center}
\begin{tabular}{l c l l l}
\hline
Sym. & State   &  $E_r$+$\Delta_s$  	&  $\Gamma_s$  &  $\tau$ [fs] \\
\hline
$^1S^e$ 	& $_2(1,0)_2^+$         &    -0.777533			     	&  0.5051[-2]			&   4.79  \\			
	        	& \cite{Chen:1997}	  &    -0.77787			     	&  0.453[-2]			&   	       \\
                 	& $_2(-1,0)_2^+$  	  &    -0.619822				&  0.2419[-3]			&  100.0 \\
		&\cite{Chen:1997}	  &    -0.62181				& 0.2178[-3]			&	      \\
		& $_2(1,0)_3^+$  	  &	-0.589631				& 0.1507[-2]			&  16.1  \\
		& \cite{Chen:1997}	  &	-0.589896				& 0.137[-2]			&            \\
		& $_2(-1,0)_3^+$	  &	-0.547755				& 0.9895[-4]			&   244  \\
		& \cite{Chen:1997}	  &  	-0.548070				& 0.775[-4]			&            \\
\hline         		 
$^1P^o$	& $_2(0,1)_2^+$	  &	-0.692642				& 0.1392[-2]			&  17.4  \\
		& \cite{Chen:1997}	  &	-0.693069				& 0.1372[-2]			&	      \\
		& $_2(1,0)_3^-$	  &	-0.597065				& 0.4069[-5]			&  5945 \\		 	
		&\cite{Chen:1997}	  &	-0.597074				& 0.384[-5]			&		\\
		&$_2(0,1)_3^+$	  &	-0.563721				& 0.2956[-3]			&  81.8	\\
		&\cite{Chen:1997}	  &	-0.564074				& 0.2998[-3]			&		\\
		&$_2(-1,0)_3^0$         &	-0.547031				& 0.2358[-7]			& $>$ 1 ns \\
		&\cite{Chen:1997}	  &	-0.547087				& 0.15[-7]				&		\\
\hline						 
$^1D^e$	& $_2(1,0)_2^+$	&	-0.701512				& 0.2528[-2]			& 9.56  \\	
		&\cite{Chen:1997}	&    	-0.70183				& 0.236[-2]			& 		\\ 
		& $_2(1,0)_3^+$	&    	-0.569448				& 0.6052[-3]			& 39.9   \\  
		& \cite{Chen:1997}	&	-0.569193				& 0.560[-3] 			&		\\
		& $_2(0,1)_3^0$	& 	-0.556317				& 0.2015[-4]			& 1200	\\
		& \cite{Chen:1997}	&	-0.556417				& 0.201[-4]			&		\\
		& $_2(1,0)_4^+$	&	-0.536575				& 0.2485[-3]			& 97.3	\\
		& \cite{Chen:1997}	&	-0.536715				& 0.234[-3]			& 		\\
\hline
\end{tabular}
\end{center}
\end{table}  

\subsection{$\cal{P}$-subspace.}
Since the illustration of the method is restricted to one- and two-photon ionization below the $N$=2 threshold in He atom, we only deal with the
simple case of single-channel continua. The multichannel case for processes above $N$=2 within the Feshbach formalism has also been
developed \cite{Sanchez:1995a,Bachau:2001,Sanchez:1995b,Martin:1993}.  A set of discretized nonresonant continuum states $\mathcal{P}\Psi^0_i$ with energies $E_i$ and angular momentum $L$ are obtained by diagonalizing Eq. (\ref{eq:Pspace}) for the $\mathcal{P}$ space using a CI expansion $\mathcal{P}\Psi^0_i=\sum_{j=1}^N  C^i_{j} \mathcal{A} (\phi_{1s}
\cdot \phi_{j \ell=L} )$ where one of the electrons is frozen in the $\phi_{1s}$ orbital (static exchange approximation) and the other $\phi_{j \ell}$ 
pertains to the subset of $N$ hydrogenic orbitals generated with $M$ B-splines with angular momentum $\ell$=$L$. Therefore, the wave function for the ejected electron corresponding to the state with discretized energy $E_i$$>$-2.0 (first ionization threshold) can be built using the CI expansion coefficients \cite{Chang:1991}, i.e.,  $\psi_i ({\bf r})$=$\sum_{j=1}^N C^i_{j} \phi_{j \ell}(\bf{r})$. This discretized continuum wavefunction is normalized to unity since it comes from a diagonalization procedure.  In order to
renormalize it to the correct Dirac delta, it suffices to multiply with the density of states factor, i.e., $\psi_{E_i}$=$[\rho(E_i)]^{1/2}$$\psi_i$, computed with a two-point formula from the set of discretized continuum energies, $\rho(E_i)= 2/(E_{i+1}-E_{i-1})$ \cite{Macias:1987}. The function $r \cdot \psi_{E_i}$ contains the effect of the Coulomb and the short-range potentials, and its radial part  behaves asymptotically as   
$\sqrt{ 2/ \pi k_{E_i} } \left[ \mathcal{F}_{k \ell} (k_{E_i}, r) \cos \delta_{E_i} + \mathcal{G}_{k\ell} (k_{E_i},r) \sin \delta_{E_i}  \right]$ \cite{Friedrich:1991}, where   $\mathcal{F}_{k \ell}$  and $\mathcal{G}_{k\ell}$ correspond to the regular and irregular Coulomb functions, respectively, and
$\delta_{E_i}$ is the scattering phase shift against the free Coulomb wave. By a least-square fitting procedure the computed function $r \cdot \psi_{E_i}$ can be adjusted to the analytical asymptotic form inside an interval [$r_{asym}, L$] in the outer part of the box of size $L$ (covering at least two wavelengths), from which the corresponding phase shifts $\delta_{E_i}$ can be obtained. 

The knowledge of these phase shifts is useful not only to compute differential cross sections, but also to provide a prescription to adjust the discretized continuum energies $\{E_i\}^N_{i=1}$ to any desired energy value. For instance, to compute the widths with Eq. (\ref{eq:width}), the nonresonant continuum state $\mathcal{P}\Psi^0$ must be degenerate with the resonant energy $E_s$. By diagonalization of the $\mathcal{P}$ eigenvalue problem, none of the $E_i$ energies coincide
with $E_s$, but $E_s$ interpolates two neighboring discretized continuum energies, $E_i < E_s < E_{i+1}$ and so do the phase shifts 
$\delta_{E_i} < \delta_{E_s} < \delta_{E_{i+1}}$. Then, the interpolated $\delta_{E_s}$ is introduced in the asymptotic radial formula given above. Every single-electron radial continuum function obtained by diagonalization inside the box satisfy the boundary condition  $r\cdot\psi_{E_i}(r$=$L)$=0, but the asymptotic analytical form with the interpolated phase shift $\delta_E$ does not. Instead, it shows a node at $r_0$ very close to the edge of the box $L$. This means that in order to
get one of the eigenvalues $E_i$ be equal to $E$=$E_s$ the box length should be replaced by $r_0$. To avoid changes in the
box length and therefore in the B-splines basis itself, one may keep the same box length $L$, but with the addition of a step potential 
$V_0 \Theta(r-r_0)$ ($\Theta$ is the Heaviside function and $V_0$ is a large real number) in the eigenproblem for the atomic orbitals.
This prescription guarantees that at least one of the eigenvalues $E_i$ is nearly degenerate with the selected $E$ value.
This general procedure can be applied to the computation of resonance widths as well as to produce a new set  of 
discretized continuum states $\{\epsilon_i\}^{M>N}_{i=1}$ with an energy spacing selected at will, which may be useful, for example, to increase the density of states if required.

In this work we have computed the nonresonant continuum states for $^1S^e$, $^1P^o$ and $^1D^e$ ($L$=0,1,2) with CI ($\phi_{1s} \phi_{j \ell=L}$) configurations, where $j$ runs up to 170 orbitals for each $\ell$=$L$ angular momentum. At least, 75 continuum states are obtained lying 
above the first ionization threshold and below the second one, the energy region where the lowest doubly excited states of $\mathcal{Q}$-space
are located.

\subsection{Time-dependent calculations}
Once the static calculations for the $\mathcal{Q}$ and $\mathcal{P}$ space are performed, the energies, dipolar couplings and the $\mathcal{Q}H\mathcal{P}$ electrostatic couplings must be introduced in Eq. (\ref{eq:coupled}), a system of differential coupled equations which is solved using a sixth-order Runge-Kutta integrator. Long time propagations (even beyond the laser pulse duration) require large radial boxes and 
energy spacings in the discretized nonresonant continuum smaller than the spectral width of the pulse (approximately given by $\Delta \omega$=
$4\pi/T$) and at least smaller than the largest resonance widths $\Gamma_s$. With only 75 continuum states between the first and the second
ionization threshold, the density of states is not large enough, but one may create {\em a larger box} by simply interpolating the dipolar
and $\mathcal{Q}H\mathcal{P}$ couplings involving the continuum with energies $\{E_i\}_{i=1}^N$. A new much denser set of continuum energies 
$\{E'_i\}_{i=1}^M$ can be generated following a quadratic formula $E'_i$=$\varepsilon_0$+$\alpha i^2$, with $\alpha$=$(\varepsilon_1-\varepsilon_0)/M^2$ where $\varepsilon_0$ and $\varepsilon_1$ correspond to the energies of the first and second ionization thresholds, and $M$ is the number of interpolating grid points. In our calculations we have used up to 2000 interpolating points for the dipolar and 
$\mathcal{Q}H\mathcal{P}$ couplings. This method does not fully replace the consistency of using larger boxes but for the purposes of our calculations it serves in practice to 
avoid unphysical spreading of the wavepacket and undesired reflections from the box boundary, for the propagation times used in this work.
For the results reported in the next section we always include all the computed bound and resonant states, and the set of {\em interpolated} continuum states for three symmetries, $^1S^e$, $^1P^o$ and $^1D^e$, a minimal angular basis for two-photon transitions.

\section{Results.}

\begin{figure}[t]
\centering
\includegraphics[scale=0.6]{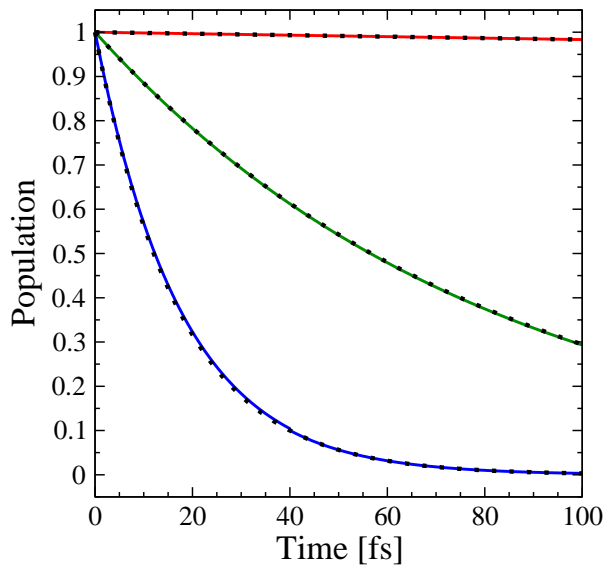}
\caption{\label{FigureI} (color online) Time decay of the three lowest $^1P^o$ resonances in He, $_2(0,1)_2^+$ (blue), $_2(1,0)_3^-$ (red) and
$_2(0,1)_3^+$ (green) quoted in table \ref{TableI}, obtained
by solving the field-free TDSE with only $\mathcal{Q}H\mathcal{P}$ couplings, for a propagation time of $t$=100 fs, and compared to the 
exponential decay formula $P(t)$=$P(0) e^{-\Gamma_r t}$ (dotted lines), where $\Gamma_r$ values are taken from table \ref{TableI}.}
\end{figure}

To illustrate the performance of the time-dependent Feshbach method in He, we restrict ourselves to the study of one- and two-photon
ionization with laser intensities corresponding to the perturbative regime, and moderately long femtosecond pulses.

\subsection{One-photon ionization from the $^1S^e$ ground state}
First, in order to test the adequacy and effectiveness of the computed $\mathcal{Q}H\mathcal{P}$ couplings, responsible for the decay of the doubly excited states into the continuum, we consider field-free time propagations of the TDSE up to 100 fs, in which the initial state corresponds to a given fully populated  resonance, i.e., $C_r$($t$=0)=1. This should represent a non stationary evolution, and the expected time-dependent probability must follow an exponential decay $P_r(t)$=$P_r(0) e^{-\Gamma_r t}$, where $\Gamma_r$ corresponds to the resonance width, computed with the
Feshbach stationary method (see figure \ref{FigureI}).
 
 \begin{figure}[t]
\centering
\includegraphics[scale=0.6]{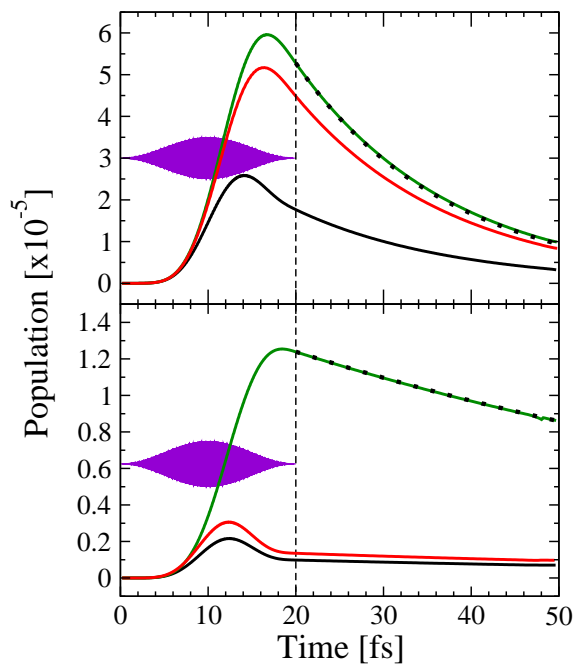}
\caption{\label{FigureII} (color online) Time resolved photoexcitation and decay of the two fastest decaying $^1P^o$ resonances in 
He, $_2(0,1)_2^+$ (upper panel) and $_2(0,1)_3^+$ (lower panel) in the one-photon ionization process from the ground state, obtained by solving the TDSE for He subject 
to a laser field with $I$=10$^{10}$ W/cm$^2$ and duration $T$=20 fs, with a total propagation time $t$= 100 fs. The color scheme corresponds 
to computations with central frequencies $\omega$ corresponding to different small positive (red) and negative (black) detunings, the line in green with the largest population corresponds to zero detuning. The dotted lines indicate the exponential decay according to $P(t)$=$P(T)e^{-\Gamma (t-T)}$. }
\end{figure}

According to table \ref{TableI} only the first and the third lowest $^1P^o$ doubly excited states have lifetimes under 100 fs, and therefore
within this limit of time propagation only features corresponding to these two resonances would be noticeable in the photoionization spectra.
One-photon ionization cross section from the ground state $^1S^e + \omega$$\to$$^1P^o$ is computed by solving the TDSE with both dipolar and $\mathcal{Q}H\mathcal{P}$ couplings included,  for a laser pulse with intensity $I$=10$^{10}$ W/cm$^2$, duration $T$=20 fs and propagation time up to $t$=100 fs, and using a grid of photon energies 24.5 eV $< E_\omega <$ 68 eV. For each photon energy, the total 
ionization probability $P_{TDSE} (t)$ is collected at times when the amplitude of the vector potential $\mathbf{A}(t)$ vanishes (the nodes of the 
function chosen for $\mathbf{A}(t)$), to comply with a proper field-free projection {\em during} the laser pulse. Ionization probabilities during and after the pulse duration come from interfering amplitudes due to the laser direct ionization and to the time delayed non stationary decay of resonances, and this interference is known to produce the typical Fano profiles in the photoabsorption spectra. When the laser field is present, and for photon energies close to the resonant condition $E_{\omega}=E_r -E(1 ^1S^e)$, the doubly excited states are populated from the ground state from the
onset of the laser pulse, reaching the maximum population in the second half of the pulse duration, from which the resonances manifestly
decay into the continuum with the exponential law. The time-resolved photoabsorption and time-delayed decay of the two most representative (fast-decaying)  $^1P^o$ resonances is pictured in figure \ref{FigureII} for different photon detunings. 

\begin{figure}[t]
\centering
\includegraphics[scale=0.7]{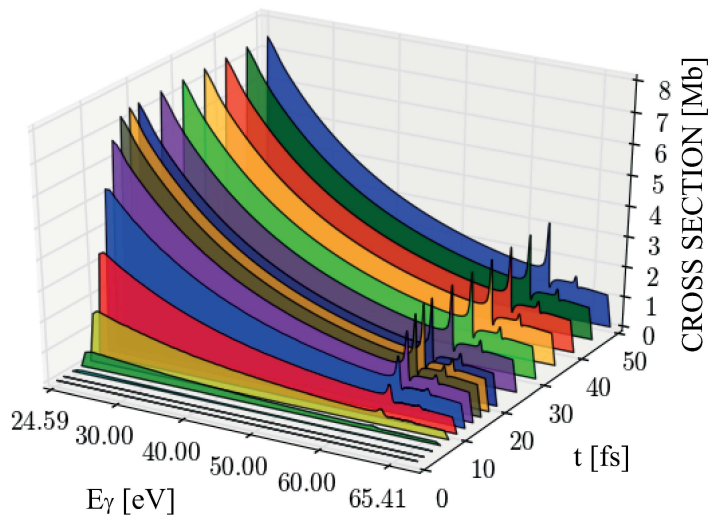}
\caption{\label{FigureIII} (color online) Temporal built-up of the one-photon ionization cross section from the ground state of He to the $^1P^o$
continuum between the first He$^+$($N$=1) and second He$^+$($N$=2) ionization threshold. Cross sections according to Eq. (\ref{eq:crosssection}) are given in Mbarn, photon energies in eV and propagation time in femtoseconds. Length and velocity gauge results are indistinguishable in the figure.}
\end{figure}

Fano interferences take place during the laser pulse but only at very long propagation times (longer than the lifetime of the slowest decaying resonance in the Rydberg series)  the Fano profile reaches its asymptotic form for all resonances. Nevertheless, the onset
and time evolution of the Fano profile corresponding to the fastest decaying $^1P^o$ resonances can be studied within the practical limitations
of the present application due to the limited chosen basis and box size. In figure \ref{FigureIII} we plot the time dependent built-up of the 
photoionization cross section between the first and the second ionization thresholds for the same laser pulse described above. In fact, a 
laser pulse with duration $T$=20 fs allows to visualize the formation of the nonresonant background as well as the two major resonant peaks present 
in the time-independent perturbative spectrum. After the laser pulse duration there are slight changes in the total ionization 
and after $t$=50 fs the behavior is almost stationary.  From this figure we may appreciate that the resonant peaks appear very soon after the 
pulse is switched on, but their profile is almost symmetrical. Only in the second half of the laser pulse, the resonant states become 
noticeably populated, subsequently decaying and the interfering with the direct ionization, which produces ultimately the final asymmetry in the 
profiles. 

\begin{figure}[t]
\centering
\includegraphics[scale=0.6]{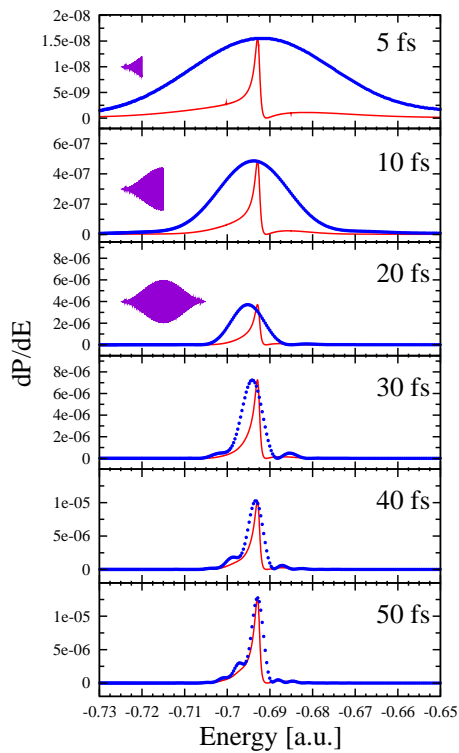}
\caption{\label{FigureIV} (color online) Time-resolved formation of the Fano profile in the differential photoionization probability $dP/dE$ in the energy region of the lowest He $^1P^o$ resonant state, using a laser pulse with central frequency  $\omega$=2.21 a.u., intensity $I$=10$^{10}$ W/cm$^2$ and duration $T$=20 fs (the time evolution of the laser pulse is also drawn). Snapshots taken at 5, 10, 20, 30 and 50 fs. The red solid line
corresponds to the result obtained with an analytical model \cite{Mercouris:2007}  (see text). The blue dots are calculated with the 
time-dependent Feshbach formalism.}
\end{figure}

\begin{figure}[t]
\centering
\includegraphics[scale=0.6]{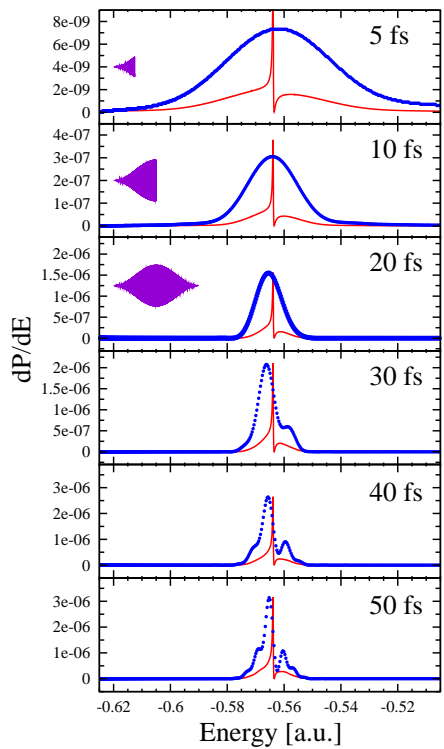}
\caption{\label{FigureV} (color online) The same as in figure \ref{FigureIV} but for the third lowest He $^1P^o$ resonant state, then using a laser pulse with central frequency  $\omega$=2.34 a.u.}
\end{figure}

\begin{figure}[t]
\centering
\includegraphics[scale=0.6]{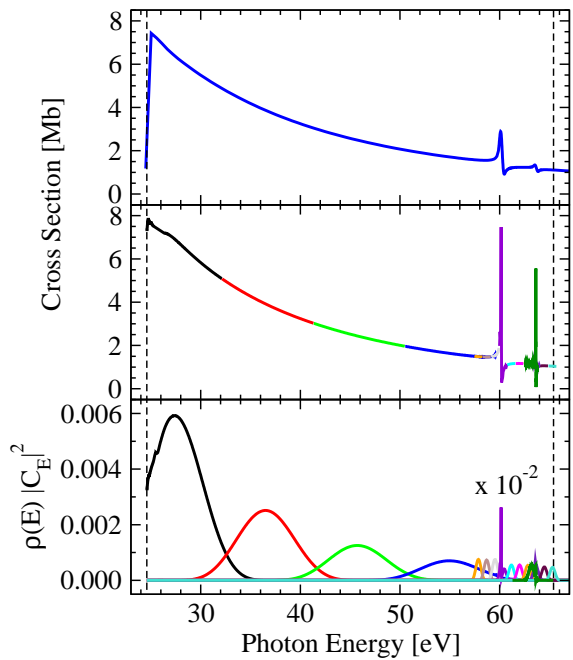}
\caption{\label{FigureVI} (color online) Top panel: One-photon single ionization cross section [in Megabarns], against the photon energy [in eV]
of He, obtained  solving the Feshbach TDSE with a laser pulse of duration $T$=20 fs and intensity $I$=10$^{12}$ W/cm$^2$, with a propagation
time of $t$=100 fs. Middle panel: One-photon single ionization cross sections, extracted from the differential photoionization probabilities 
$dP/dE$=$\rho(E) |C_E|^2$ plotted in the bottom panel, obtained using a laser pulse with shorter duration $T$=0.9 fs and intensity $I$=10$^{12}$ W/cm$^2$ and central frequencies $\omega$=27.57 eV (black), 36.76 eV (red), 45.95 eV (green), 55.14 eV  (blue). Sharper signals for the
resonant peaks are obtained extracting cross sections from differential probabilities (scaled with a factor 10$^{-2}$ in the figure) 
calculated with longer pulse duration $T$=10 fs for
photon energies between $\sim$57 and $\sim$65 eV. Vertical  dashed lines indicate ionization thresholds He$^+$($N$=1) and He$^+$($N$=2). All reported results in velocity gauge. }
\end{figure}

The time dependent formation of the Fano profile can also be analyzed from the photoionization probabilities, differential in energy, $dP(E,t)/dE$.
To obtain reliable differential probabilities in the continuum a rather good density of continuum states close to the resonance positions is required,
and they are simply computed with the continuum expansion coefficients, i.e.,  $dP(E,t)/dE=\rho(E) |C_E(t)|^2$, where $\rho(E)$ is the density of states. Nicolaides {\em et al} \cite{Mercouris:2007} have also recently studied the time dependent formation of the profile of the lowest He $2s2p$ (or  $_2(0,1)_2^+$) $^1P^o$ resonant state, induced by a short laser pulse. Apart from their {\em ab initio} calculation using an spectral method 
called {\em state specific expansion approach}, they propose a simplified theoretical model
to account for the profile formation. In their simplified model the transition amplitude to the continuum is given by
\begin{eqnarray} \label{eq:NicolaidesA}
A_E(t) & =  C_E e^{-i E t} G(E,t) \left[ q + \frac{E- (\mathcal{E}_r + \Delta_r)}{\Gamma_r} \right]  \nonumber \\
             & -C_E \frac{q-i}{2\pi i} e^{-i ( \mathcal{E}_r + \Delta_r -i\Gamma_r  )  } G(\mathcal{E}_r+\Delta_r -i\Gamma_r, t),
\end{eqnarray}
where $\{\mathcal{E}_r, \Delta_r, \Gamma_r, q \}$ are the set of resonance parameters corresponding to uncorrected position, energy shift, width and $q$ Fano shape parameter, respectively, the coefficient $C_E$=$\Gamma_r d_{gE} / \sqrt{[E-(\mathcal{E}_r+ \Gamma_r)  ]^2 + \Gamma_r^2 } $, with 
$d_{gE}$ being the dipolar coupling value between the ground state and the scattering state with energy $E$. The $G(E,t)$ function contains the
dependence on the laser pulse and it is basically the Fourier transform of the electric field of the laser pulse, i.e., $G(E,t)$=$-i \int_0^t dt' e^{-i (E_g-E) t'} \mathbf{E}(t) \hat{{\bf e}}_z$. We compare our {\em ab initio} results with this simplified model, for the same laser parameters described above with a central frequency $\omega$=2.211 a.u., $I$=10$^{10}$ W/cm$^2$ and duration $T$=20 fs and the resonance parameters
quoted in table \ref{TableI}, with $q$=-2.8 and $d_{gE}$=0.48185. We have computed {\em ab initio} differential ionization probabilities for 
continuum energies close to the lowest $^1P^o$ resonance, -0.75 $<$ $E$[a.u.] $<$ -0.63, and they are plotted in figure \ref{FigureIV}. 
We notice that the analytical model reaches the asymmetric profile much more rapidly than the {\em ab initio} method, the latter showing
the effect of laser spectral width at short times, in the sense that the He atom is not yet {\em aware} of the total pulse duration. The Feshbach results only converge to the expected asymptotic result of the model for times much longer than the pulse duration, with small transient oscillations due to interferences associated to the still active resonance decay.

The same comparison is done with the third lowest $^1P^o$ resonance in figure \ref{FigureV} in the energy range -0.62 $<$ $E$[a.u.] $<$ -0.5, now with a central frequency $\omega$=2.34 a.u. a dipolar coupling value $d_{gE}$=0.439891 a.u. and a Fano parameter $q$=-2.5, and the resonance parameters quoted in table \ref{TableI}.  Again, the Feshbach result nicely shows the trend of convergence to the analytical result of the model for
propagation times much larger than the pulse duration. Of course, our {\em ab initio} simulations in the present illustrations are constrained by our box length $L$=150 a.u., and therefore the total propagation time is also limited. In fact, this drawback in time dependent computations has been
recently overcome. Usually, autoionizing lifetimes may be much longer than the pulse duration and the required field-free integration times are
prohibitive in order to obtain asymptotic ionization probabilities in the neighborhood of all resonances conforming a Rydberg series. In a recent
series of papers by Palacios {\em et al} \cite{Palacios:2007,Palacios:2008} it is shown that multiphoton ionization cross sections can be retrieved
from differential amplitudes obtained with very short laser pulses, the asymptotic amplitudes being extracted by solving a driven equation
within the exterior complex scaling approach. This method is in practice equivalent to propagate the time-dependent wave-packet for an infinite
time after the pulse duration. This methodology could be eventually incorporated to our Feshbach time-dependent method, but this will not be done here. Instead, we use their procedure to obtain photoionization cross sections (for which one assumes low intensities and laser pulses with infinite duration) from differential ionization amplitudes obtained with short laser pulses, thanks to the factorability of the time dependence in first-order perturbation theory, in the form of the Fourier transform of the pulse.

For instance, using expression (17) in Ref. \cite{Palacios:2008}, modified to be used with our differential probability amplitudes in the energy scale,
we can reproduce both the nonresonant background and the resonant sharp peaks in one-photon ionization cross section using transition 
probabilites obtained within the spectral bandwidth of shorter laser pulses with durations $T$=0.9 fs and $T$=10 fs, the latter to improve the resolution in the resonance region (see figure \ref{FigureVI}). Again, since we are not using truly asymptotic amplitudes ($t\to\infty$) we cannot obtain the high resolution in Fig. 2 of Ref. \cite{Palacios:2008} for the Rydberg series
of $^1P^o$ resonances. 

\subsection{One-photon ionization from the lowest $1s2p$ $^1P^o$ excited state.}

\begin{figure}[h]
\centering
\includegraphics[scale=0.5]{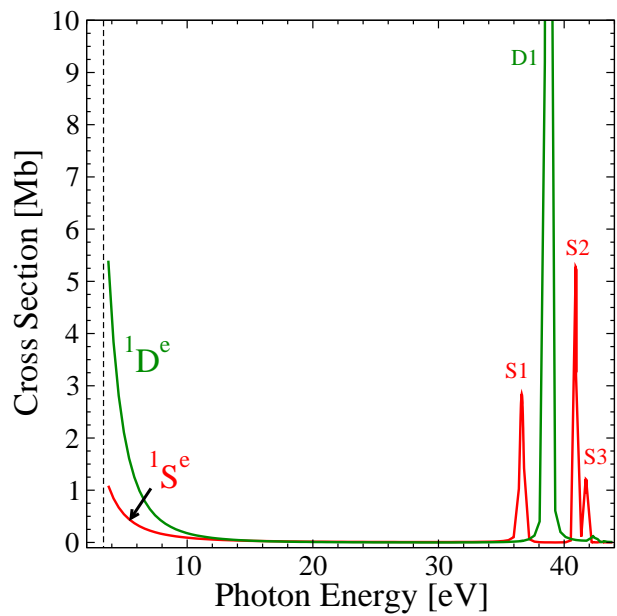}
\caption{\label{FigureVII} (color online) One-photon ionization cross section from the $1s2p$ $^1P^o$ excited state to partial final contributions 
$^1S^e$ (red) and $^1D^e$ (green), obtained from our Feshbach time-dependent method, using a laser pulse with intensity $I$=10$^{10}$ W/cm$^2$ and duration $T$=20 fs. Peaks denoted with $S$ correspond to $^1S^e$ doubly excited states and that with $D$ correspond to the lowest
$^1D^e$ doubly excited state, all quoted in Table \ref{TableI}. Results in velocity gauge are only included.}
\end{figure}

In order to reach also the lowest $^1S^e$ and $^1D^e$ resonances, one may check out
the one-photon ionization from the $1s2p$ $^1P^o$  state. According to table \ref{TableI} the lowest resonances in these two symmetries
decay even faster than those in $^1P^o$. The cross section obtained for a laser pulse of duration $T$=20 fs and total propagation $t$=100 fs is plotted in figure \ref{FigureVII}. The observed peaks in the spectrum correspond to the three lowest $^1S^e$ doubly excited states quoted in 
table \ref{TableI}
and the lowest $^1D^e$ resonant state. Our cross sections compare well with the perturbative stationary result, computed with a Multiconfigurational
Hartree-Fock (MCHF) method and B-splines \cite{Froese-Fischer:1990}. The $1s2p$ $^1P^o$ excited state is closer to the upper ionization than
the ground $^1S^e$ state. A photon energy of $\sim$21.2 eV is able to photoionize the He atom, but this energy is also resonant with the
ground $^1S^e$ state. In this case the laser field can couple different one- and three-photon processes to yield $^1S^e$ and $^1D^e$ 
enhanced ionization probabilities, along with Rabi oscillations between the initial state $^1P^o$ and the ground state $^1S^e$, 
nonlinear phenomena hardly seen at this perturbative low laser intensities, but expected to play a major role at higher intensities. 

\subsection{Two-photon ionization from the $^1S^e$ ground state.}

\begin{figure}[t]
\centering
\includegraphics[scale=0.8]{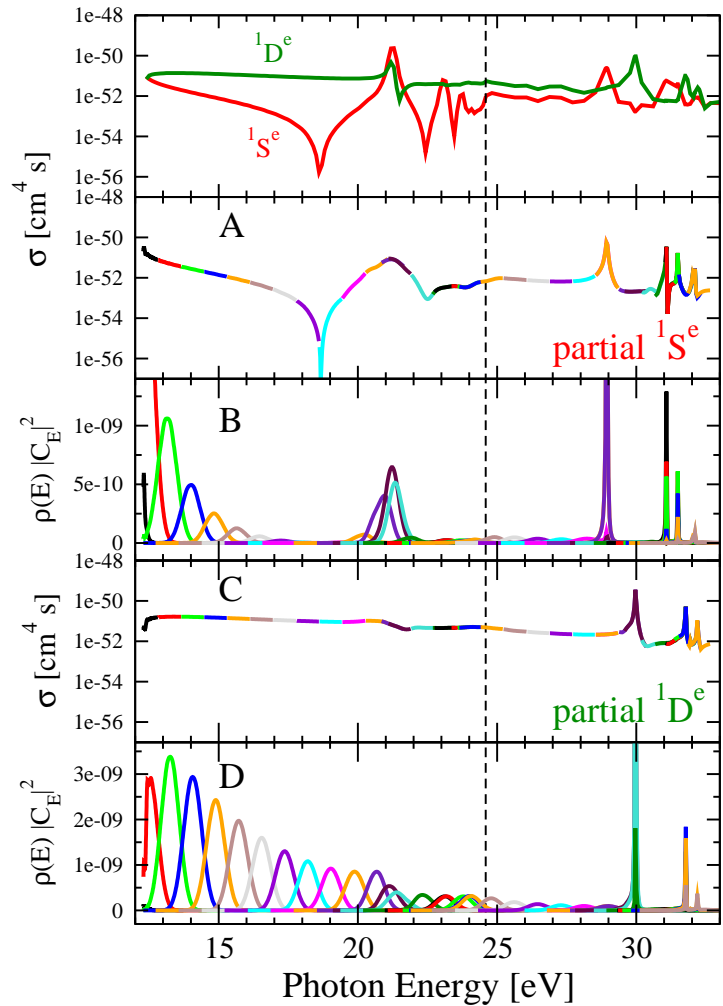}
\caption{\label{FigureVIII} (color online) Two-photon ionization cross section from the $^1S^e$ ground state in He.
Top panel: $^1S^e$ (red) and $^1D^e$ (green) components of the cross section, computed by solving the Feshbach-TDSE with
pulses of intensity $I$=10$^{10}$ W/cm$^2$ and duration $T$=20 fs. Panel A: $^1S^e$ component of the two-photon
cross section applying the piecewise renormalization procedure of Ref. \cite{Palacios:2008} and using the computed 
differential probabilities included in panel B, obtained with laser pulses with duration $T$=5 fs and the same intensity.
Every segment in the reconstructed cross section is shown with a different color scheme. Panel C an D: The same as
Panel A and B, but for the $^1D^e$ component of the total cross section. The dashed vertical line indicates the above threshold
ionization threshold.}
\end{figure}

For the sake of completeness of this dynamical study, we also compute the two-photon ionization process from the ground state, in which
the $^1S^e$ and $^1D^e$ doubly excited states can be also excited with the absorption of a first photon into the intermediate $^1P^o$ 
continuum (above threshold ionization). Here, continuum-continuum dipolar matrix elements are explicitly included, and in this case it is known 
that calculations for above threshold ionization in the velocity gauge converge faster that in length gauge \cite{Cormier:1996}, 
so our results in figure \ref{FigureVIII} are given only in the velocity gauge. A first computation (top panel in figure \ref{FigureVIII}) is carried out with a laser pulse with intensity 
$I$=10$^{10}$ W/cm$^2$ and a pulse duration $T$=20 fs, with a total integration time $t$=100 fs and we plot the separate $^1S^e$ and 
$^1D^e$ partial contributions. The peaks corresponding to resonant intermediate $^1P^o$ bound states in the photon energy region
[21.2,24.6] eV are reasonably well resolved. In the above threshold ionization region $\omega >$ 24.6 eV, the two major features correspond
to peaks generated by the $^1S^e$ ($2s^2$ or $_2(1,0)^+_2$) lowest resonant state and  the $^1D^e$ ($2p^2$ or $_2(1,0)^+_2$) lowest state.
A better resolution for the resonant peaks in the above threshold ionization region can be achieved by extracting the cross sections from the
photoionization amplitudes obtained by a short pulse and renormalized by the two-photon shape function of the laser pulse (see Ref. 
\cite{Palacios:2008} for details).  

Adapting the expression (25) in \cite{Palacios:2008} to the energy scale and using our photoionization amplitudes
$\rho^{1/2}(E) C_E$ (which in our case are not truly asymptotic) we may reproduce the two-photon ionization cross sections piecewise, 
using an assortment of laser pulses of duration $T$=5 fs with central frequencies from 12.4 to 32.26 eV.  In panel A of 
figure \ref{FigureVIII} we show the piecewise reconstruction of the $^1S^e$ component in the total cross section, obtained from
differential ionization probabilities $\rho(E)|C_E|^2$ shown in the panel B below. Similarly, the $^1D^e$ component in panel C 
is generated following the same procedure with the differential probabilities shown in panel D.  With this {\em pulse renormalization} 
procedure (valid only in the perturbative regime), the structure of resonant peaks in the Rydberg series are more cleanly discriminated, but at 
the cost of partially loosing the structures due to one-photon absorption to intermediate $^1P^o$ bound states, which require
longer laser pulse durations to be consistently resolved \cite{Palacios:2008}.

\section{Conclusion}

In conclusion, this work describes the theoretical details and the inner workings of an {\em ab initio} time dependent Feshbach method as 
applied to the resonant photoionization of He atom using ultrashort laser pulses, below the second ionization threshold. 
Some simple illustrations of the performance of the method have been included, related to one and two-photon ionization processes, 
without further sophistications. It is assumed that there is much room for improvements, in terms of a optimized grids of B-splines, 
much larger radial boxes, larger configurational basis and partial waves for better convergence, the introduction of complex  
absorbing potentials, or even exploring other time propagators. At the present level, all computations presented here have been performed in 
simple desktop computers. In order to deal with pump-probe-like experiments where a XUV pump laser excites the resonances to be subsequently probed by an IR laser field, the multichannel extension of the method is required, but it can be implemented straightforwardly within the Feshbach formalism \cite{Bachau:2001}. Steps along these directions are under way.

\section*{Acknowledgements}

The authors acknowledge financial support from Vicerrector\'{\i}a de Investigaci\'on at the Universidad de 
Antioquia, Colombia.  C.M. Granados-Castro gratefully acknowledges the COOPEN program for a semester stay at Physik Department, Technische Universit\"at M\"unchen (TUM). The authors thank the hospitality extended to them by Harald Friedrich and Javier Madro\~nero at TUM and for helpful discussions. J.L. S-V also thanks F. Mart\'{\i}n and A. Palacios for discussions and encouragement.

\end{document}